\documentclass[twocolumn,showpacs,superscriptaddress,amsmath,amssymb,aps,pra]{revtex4-1}
\usepackage{bm}
\usepackage{graphicx}
\usepackage{dcolumn}
\usepackage{amsmath,txfonts}
\usepackage{epstopdf}
\usepackage[mathlines]{lineno}
\usepackage{color}
\usepackage[colorlinks=true,linkcolor=blue,citecolor=blue]{hyperref}
\usepackage[normalem]{ulem}

\begin{document}

\title{Superradiant phase transition in cavity magnonics via Floquet engineering}

\author{Si-Yan Lin}
\affiliation{School of Physics, Hangzhou Normal University, Hangzhou 311121, China}

\author{Fei Gao}
\affiliation{School of Physics, Hangzhou Normal University, Hangzhou 311121, China}

\author{Ye-Jun Xu}
\affiliation{Interdisciplinary Research Center of Quantum and Photoelectric Information,
and Anhui Research Center of Semiconductor Industry Generic Technology, Chizhou University, Chizhou, Anhui 247000, China}

\author{Lijiong Shen}
\affiliation{School of Physics, Hangzhou Normal University, Hangzhou 311121, China}

\author{Yan Wang}
\affiliation{School of Physics, Hangzhou Normal University, Hangzhou 311121, China}

\author{Xiao-Qing Luo}
\email{xqluophys@gmail.com}
\affiliation{School of Physics and Electronic Information, Yunnan Normal University, Kunming, 650500, China}

\author{Guo-Qiang Zhang}
\email{zhangguoqiang@hznu.edu.cn}
\affiliation{School of Physics, Hangzhou Normal University, Hangzhou 311121, China}
\affiliation{Zhejiang Key Laboratory of Micro-Nano Quantum Chips and Quantum Control, Hangzhou 310027, China}

\begin{abstract}
We propose a scheme to engineer the superradiant phase transition (SPT) in cavity magnonics by periodically modulating the frequency of the magnon mode. The studied system is composed of a yttrium iron garnet (YIG) sphere positioned inside a microwave cavity, where magnons in the YIG sphere are strongly coupled to microwave photons. Under the Floquet drive, the effective frequencies of both the cavity and magnon modes can be readily controlled via the frequency and strength of Floquet field. This tunability allows the cavity magnonic system to support a rich steady-state phase diagram, featuring parity-symmetric, parity-symmetry-broken, bistable, and unstable phases. With the increase of Floquet-field strength, the system exhibit a discontinuous phase transition from the parity-symmetric phase to the parity-symmetry-broken phase at a critical threshold, accompanied by an abrupt jump of the magnon occupation from zero to a finite value. Upon further increase of Floquet-field strength, the magnon occupation declines continuously from a nonzero value back to zero, corresponding to a second-order phase transition that restores the parity-symmetric phase. Additionally, fluctuations in magnon number during the SPT process are examined. Our work establishes an alternative route to engineer the cavity-magnon SPT without relying on microwave parametric drive.
\end{abstract}

\date{\today}

\maketitle

\section{Introduction}

The last decade has witnessed the rapid development of cavity magnonics, where magnons (i.e., collective spin excitations) in ferromagnetic materials, e.g., yttrium iron garnet (YIG), are strongly coupled to photons in microwave cavities~\cite{Rameshti22,Yuan22}. Thanks to their high tunability and design flexibility, cavity magnonic systems are a versatile platform for exploring diverse novel physical phenomena, such as dissipative couplings~\cite{Harder18,Grigoryan18,Wang22,Yu19}, magnon dark modes~\cite{Zhang15,Zhang23}, exceptional points~\cite{Cao19,Zhang19,Rao24,Lambert25}, nonclassical states of magnons~\cite{Quirion20,Sun21,Xu23}, nonreciprocal microwave transmission~\cite{Wang19,Zhang20}, quantum entanglement~\cite{Li18}, and precision measurements~\cite{Trickle20,Wolski20,Zhang23PRB}. In particular, on-demand control of the cavity-magnon interaction has been experimentally realized via Floquet engineering~\cite{Xu20,Pishehvar25}, i.e., temporal modulation of system parameters by periodic drives~\cite{Eckardt17,Stehlik16,Han19,Sameti19}. This approach surpasses the conventional rotating-wave picture, allowing the cavity magnonic system to access the Floquet ultrastrong coupling. Such Floquet-engineering control enables cavity magnonic systems to explore, e.g., topological simulations~\cite{Hei24}, enhanced quantum effects~\cite{Li24,Xie23,Guan24}, generation of NOON states~\cite{Qi23}, and chiral couplings~\cite{Ren22}. Furthermore, various intriguing nonlinear phenomena induced by magnon Kerr effect have also been investigated in cavity magnonic systems. The magnon Kerr effect, originating from the magnetocrystalline anisotropy in YIG, mediates interactions among magnons~\cite{Zhang-China-19,Wang16}. Under the strong coherent drive, this Kerr nonlinearity can give rise to bistability and multistability of cavity magnon polaritons~\cite{Wang18,Shen22,Bi21,Bi24,Nair21}, high-order sidebands~\cite{Wang21,Zhao22}, nonreciprocal quantum effects~\cite{Chen23,Ahmed25,Kong24,Zhang24,Lai25}, long-distance spin-spin coupling~\cite{Xiong22}, and nonlinear spin currents~\cite{Nair20,Shen21}, among other phenomena.

Superradiant phase transition (SPT) was first proposed in the Dicke model~\cite{Hepp73,Wang73}, which describes the collective coupling of a spin ensemble to a quantized cavity field~\cite{Dicke54}. As the coupling strength increases beyond a critical threshold, the system undergoes a phase transition from the normal phase to the superradiant phase~\cite{Emary03PRL,Emary03PRE}. However, observing this original SPT experimentally remains challenging, because the required critical coupling strength is comparable to the frequencies of both the spin ensemble and the cavity mode, making it difficult to achieve. To circumvent this difficulty, the nonequilibrium SPT (i.e., the simulated SPT) has been proposed and demonstrated, where an effective Dicke-type interaction is engineered in driven cavity QED systems (see, e.g., Refs.~\cite{Dimer07,Baumann10,Zou14,Baden14,Zhu20,Zhang21Science,Wu23,Huang23,Zheng23,Zhu24,Zhao25}). For example, Ref.~\cite{Zhang21} proposed a scheme to observe the nonequilibrium SPT in a cavity magnonic system, leveraging the cooperative effects of the magnon Kerr nonlinearity and the parametric drive. Subsequently, Refs.~\cite{Qin22} and \cite{Xu24} have further explored controllable and nonreciprocal versions of this SPT in a double-cavity magnonic system and in a spinning resonator setup, respectively. Notably, a critical and experimentally demanding prerequisite for all these schemes~\cite{Zhang21,Qin22,Xu24} is the necessity of applying parametric drive, which imposes stringent conditions and complicates practical implementation. In this context, developing alternative schemes that do not rely on parametric drive becomes particularly urgent and necessary.

This work theoretically introduces a new approach that enables the observation of the SPT in cavity magnonics without employing microwave parametric drive. We consider a cavity magnonic system in which a magnon mode with Kerr nonlinearity in a YIG sphere is strongly coupled to a microwave cavity~\cite{Wang16,Wang18}. Our approach incorporates Floquet engineering into the cavity magnonic setup~\cite{Xu20,Pishehvar25}, allowing flexible tuning of the cavity-magnon coupling. By applying two unitary transformations, the time-dependent terms in the system Hamiltonian can be eliminated, yielding an anisotropically coupled two-harmonic-oscillator model. Using quantum Langevin equations, we investigate the steady state of the system and map out its steady-state phase diagram. The phase diagram reveals four distinct regions: parity-symmetric phase, parity-symmetry-broken phase, bistable phase and unstable phase. As the drive strength of Floquet field increases, the system first undergoes a first-order SPT from the parity-symmetric phase (characterized by no macroscopic magnon occupation) to the parity-symmetry-broken phase, which exhibits macroscopic magnon occupation. Upon further increasing the Floquet-field strength, the system returns to the parity-symmetric phase via a second-order phase transition. Near the critical points of these transitions, magnon number fluctuations diverge, which is a hallmark signature of SPT.

In contrast to earlier proposals for the cavity-magnon SPT that rely on parametric drive~\cite{Zhang21,Qin22,Xu24}, our scheme is based on the cooperative interplay between magnon Kerr nonlinearity and Floquet modulation. Although a Josephson parametric amplifier can, in principle, generate the required microwave parametric drive~\cite{Yamamoto08,Zhong13,Lin13}, its experimental integration into a cavity magnonic system remains to be demonstrated. In comparison, Floquet modulation of the cavity-magnon coupling has been realized in recent experiments~\cite{Xu20,Pishehvar25}. Moreover, implementing the theoretical schemes of Refs.~\cite{Zhang21,Qin22,Xu24} experimentally requires a microwave cavity with a specific coplanar waveguide geometry. Our scheme removes this constraint and is compatible with a wider range of microwave cavity geometries, including three-dimensional cavities~\cite{Wang16,Wang18}, superconducting coplanar waveguide resonators~\cite{Huebl13,Morris17,Hou19,Song25}, split-ring resonators~\cite{Qian20,Qian24}, and coaxial-like cavities~\cite{Haigh15,Bourhill16}. This work therefore presents an alternative approach to engineer the SPT in cavity magnonics via a distinct physical mechanism.

\begin{figure}
\includegraphics[width=0.48\textwidth]{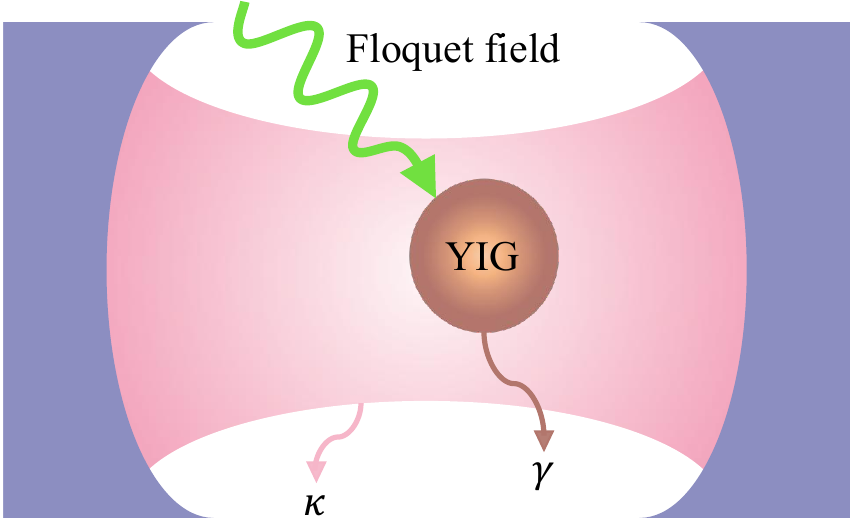}
\caption{Schematic diagram of the Floquet-driven cavity magnonic system. The system is composed of a microwave cavity and a YIG sphere driven by a Floquet field, where the magnon mode is strongly coupled to the cavity mode.}
\label{fig1}
\end{figure}

\section{Effective Hamiltonian of the cavity magnonic system}\label{EH}

As schematically depicted in Fig.~\ref{fig1}, the cavity magnonic system consists of a microwave cavity and a YIG sphere, where the YIG sphere is driven by a Floquet field. Here we assumed that the YIG sphere is uniformly magnetized to saturation by a bias magnetic field, and only the uniform precession magnon mode (i.e., Kittel mode) is strongly coupled to the microwave cavity mode. Due to the magnetocrystalline anisotropy of the YIG, there exists the interaction among magnons, i.e., magnon Kerr effect~\cite{Wang16,Wang18,Zhang-China-19}. Including both the magnon Kerr effect and the Floquet drive, the total Hamiltonian of the cavity magnonic system reads
\begin{eqnarray}\label{Hamiltonian}
\widetilde{H}&=&\tilde{\omega}_{c} a^{\dag} a +\tilde{\omega}_{m}b^{\dag}b + \frac{K}{2}b^{\dag}b^{\dag}bb + g_m(a^{\dag}+a)(b^{\dag}+b)\nonumber\\
              & &+\Omega b^{\dag}b\cos(\omega_{D}t),
\end{eqnarray}
where $a^{\dag}$ and $a$ ($b^{\dag}$ and $b$) are the creation and annihilation operators of the cavity mode (magnon mode) with resonance frequency $\tilde{\omega}_c$ ($\tilde{\omega}_m$), $K$ is the nonlinear coefficient of the magnon Kerr effect, and $g_m$ denotes the coupling strength between the cavity mode and the magnon mode. For the bias magnetic field aligning the crystalline axis $[100]$, the Kerr coefficient is positive (i.e., $K>0$). The last term in Eq.~(\ref{Hamiltonian}) describes the Floquet drive~\cite{Xu20,Pishehvar25}, where $\Omega$ and $\omega_{D}$ refer to the strength and the frequency of Floquet field. In our paper, we focus on the case of $\{\tilde{\omega}_c, \,\tilde{\omega}_m, \,\omega_D\}_{\rm min} \gg g_m$.

To derive an effective Hamiltonian in the deep-strong coupling regime, we introduce the unitary transformation
\begin{eqnarray}
U_{1}(t)=\exp\left(-i\int_{0}^{t} \left\{\tilde{\omega}_{c} a^{\dag} a +\left[\tilde{\omega}_{m}
                   +\Omega\cos(\omega_{D}\tau)\right]b^{\dag}b\right\}d\tau\right)~~~~
\end{eqnarray}
to define a rotating frame with respect to the cavity frequency $\tilde{\omega}_c$ and the magnon frequency $\tilde{\omega}_{m}+\Omega\cos(\omega_{D}t)$. In the rotating frame, the total Hamiltonian $\widetilde{H}$ given in Eq.~(\ref{Hamiltonian}) becomes
\begin{eqnarray}\label{heff1}
H'_{\rm eff}&=&U_{1}^{\dag}(t) \widetilde{H} U_{1}(t)-iU_{1}^{\dag}(t)\frac{\partial U_{1}(t)}{\partial t}\nonumber\\
   &=&\frac{K}{2}b^{\dag}b^{\dag}bb+\sum_{n=-\infty}^{+\infty}g_m J_{n}(\xi)\Big[\big(a^{\dag}be^{-i\delta_{n,-}t}
         +ab^\dag e^{i\delta_{n,-}t}\big)\nonumber\\
   & &     +\big(a^{\dag}b^\dag e^{i\delta_{n,+}t} +ab e^{-i\delta_{n,+}t}\big)\Big],
\end{eqnarray}
where the Jacobi-Anger identity $\exp[i\xi\sin(\omega_{D}t)]=\sum_{n=-\infty}^{+\infty}J_{n}(\xi)\exp(in\omega_{D}t)$ has been used, with integer number $n$ and reduced drive amplitude $\xi=\Omega/\omega_{D}$. Here, $J_{n}(\xi)$ is the $n$th Bessel function of the first kind, and $\delta_ {n,-}=\tilde{\omega}_{m}-\tilde{\omega}_{c}+n\omega_{D}$ and $\delta_{n,+}=\tilde{\omega}_{m}+\tilde{\omega}_{c}+n\omega_{D}$ are the oscillating frequencies of the $n$th rotating-wave term and the $n$th counter-rotating term, respectively. Choosing appropriate system parameters, the $n_1$th rotating-wave term satisfies $\delta_{n_1,-} \lesssim g_m J_{n_1}(\xi)$, while all other rotating-wave terms ($n \neq n_1$) exhibit high-frequency oscillations with $\delta_{n,-} \gg g_mJ_{n}(\xi)$. Similarly, the $n_2$th counter-rotating term obeys $\delta_{n_2,+} \lesssim g_m J_{n_2}(\xi)$, whereas the remaining counter-rotating terms ($n \neq n_2$) operate in the regime of $\delta_{n,+} \gg g_m J_{n}(\xi)$. According to the rotating-wave approximation~\cite{Walls94}, we can neglect all high-frequency oscillating terms, retaining only both the $n_1$th rotating-wave term and the $n_2$th counter-rotating term. The system Hamiltonian then reduces to
\begin{eqnarray}\label{heff2}
H''_{\rm eff}&=&\frac{K}{2}b^{\dag}b^{\dag}bb
                          +\lambda_r(a^{\dag}be^{-i\delta_{n_1,-}t}+ab^\dag e^{i\delta_{n_1,-}t}) \nonumber\\
                     & &+\lambda_{\rm cr}(a^{\dag}b^\dag e^{i\delta_{n_2,+}t} +ab e^{-i\delta_{n_2,+}t}),
\end{eqnarray}
where $\lambda_{r}=g_m J_{n_{1}}(\xi)$ represents the effective coupling strength of the rotating-wave interaction, while $\lambda_{\text{cr}}=g_m J_{n_{2}}(\xi)$ corresponds to the effective coupling strength of the counter-rotating interaction.
Due to $\{|J_{n_{1}}|,\, |J_{n_{2}}|\}_{\rm max} \leq 1$ for any $n_1$ and $n_2$, the effective coupling strengths cannot exceed the original cavity-magnon coupling strength, i.e., $\{|\lambda_r|, \, |\lambda_{\rm cr}|\}_{\rm max} \leq g_m$.

\begin{figure}
\includegraphics[width=0.48\textwidth]{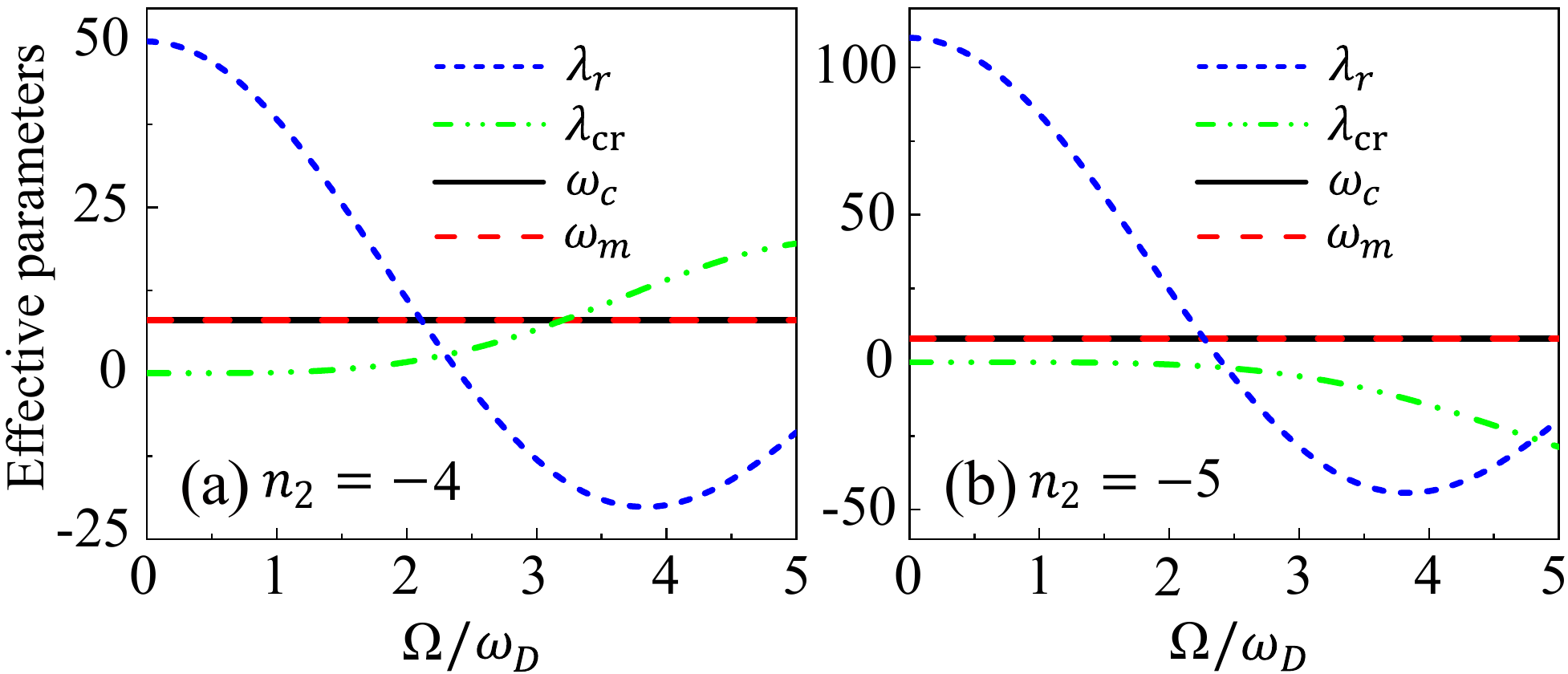}
\caption{The effective frequencies $\{\omega_c,\omega_m\}$ and coupling strengths $\{\lambda _r,\lambda _{\rm cr}\}$ versus the reduced drive strength $\Omega/\omega_D$, where $n_2 = -4$, $g_m/\kappa = 50$ in panel (a) and $n_2 = -5$, $g_m/\kappa = 110$ in panel (b). Other parameters are $\tilde{\omega}_c=\tilde{\omega}_m $, $\tilde{\omega}_c + \tilde{\omega}_m + n_2 \omega_D = 16\kappa$, $\kappa=\gamma=1$, and $n_1 = 0$~\cite{Xu20,Pishehvar25,Wang18}.}
\label{fig2}
\end{figure}

To eliminate the explicit time dependence in the effective Hamiltonian $H''_{\rm eff}$, we introduce a second unitary transformation
\begin{eqnarray}
U_2(t)=\exp\left(-i\int_{0}^{t} \left\{-\omega_{c}a^{\dag}a-\omega_{m}b^{\dag}b\right\}d\tau\right),
\end{eqnarray}
where $\omega_{c}$ ($\omega_{m}$) is the effective frequency for the cavity (magnon) mode, given by
\begin{eqnarray}
\omega_{c}=\frac{\delta_{n_{2},+}-\delta_{n_{1},-}}{2},~~~
\omega_{m}=\frac{\delta_{n_{2},+}+\delta_{n_{1},-}}{2}.
\end{eqnarray}
Then, we further rotate the effective Hamiltonian $H''_{\rm eff}$ and obtain
\begin{eqnarray}\label{Heff-2}
H_{\rm eff}&=&U_{2}^{\dag}(t)H''_{\rm eff}U_{2}(t)-iU_{2}^{\dag}(t)\frac{\partial U_{2}(t)}{\partial t}\nonumber\\
   &=&\omega_{c}a^{\dag}a+\omega_{m}b^{\dag}b+\frac{K}{2}b^{\dag}b^{\dag}bb
            +\lambda _{r}(a^{\dag}b+ab^{\dag}) \nonumber\\
   & &+\lambda_{\text{cr}}(a^{\dag}b^{\dag}+ab).
\end{eqnarray}
It possesses a conserved parity symmetry, as evidenced by the commutation relation $[H_{\rm eff}, \mathcal{P}]=0$, where the parity operator $\mathcal{P}$ is defined as~\cite{Emary03PRL,Emary03PRE}
\begin{eqnarray}\label{}
\mathcal{P}=\exp[i\pi \mathcal{N}],~~~\mathcal{N}=a^{\dag}a+b^{\dag}b.
\end{eqnarray}
Here, the operator $\mathcal{N}$ represents the total number of excitation quanta in the cavity magnonic system.  The parity operator $\mathcal{P}$ owns two eigenvalues, $+1$ (even parity) and $-1$ (odd parity), corresponding to the system states with even and odd total excitation numbers, respectively.

In our scheme, the effective frequencies $\{\omega_c, \omega_m \}$ and coupling strengths $\{\lambda _r, \, \lambda _{\rm cr}\}$ of the effective Hamiltonian $H_{\rm eff}$ in Eq.~(\ref{Heff-2}) are readily controllable via Floquet drive. In Fig.~\ref{fig2}, we plot $\omega_c$, $\omega_m$, $\lambda _r$ and $\lambda _{\rm cr}$ as functions of the dimensionless drive amplitude $\xi$. The parameters for Fig.~\ref{fig2} are set with $\tilde{\omega}_c=\tilde{\omega}_m$, $\tilde{\omega}_c + \tilde{\omega}_m + n_2 \omega_D = 16\kappa$, and $n_1 = 0$~\cite{Xu20,Pishehvar25,Wang18}, where $\kappa$ is the damping rate of cavity mode. Additionally, we take $n_2 = -4$ and $g_m/\kappa = 50$ for Fig.~\ref{fig2}(a), and $n_2 = -5$ and $g_m/\kappa = 110$ for Fig.~\ref{fig2}(b). With this parameter set, the effective frequencies of the cavity and magnon modes, $\omega_{c}/\kappa = \omega_m/\kappa = 8$, remain independent of $\xi$, whereas the normalized coupling strengths $\lambda _r/\kappa$ and $\lambda _{\text{cr}}/\kappa$ vary with $\xi$. Near $\xi = 2.2$, the magnitudes of $\lambda _r/\kappa$ and $\lambda _{\text{cr}}/\kappa$ become comparable to those of $\omega_c/\kappa$ and $\omega_m/\kappa$. This regime allows the system to exhibit SPT, as discussed in Sec.~\ref{QPT}.

\section{Steady-state solutions of the cavity magnonic system}\label{SSS}

Starting from the effective Hamiltonian in Eq.~(\ref{Heff-2}) and considering the cavity and magnon losses, the evolution of the cavity magnonic system can be described using the following quantum Langevin equations~\cite{Walls94}:
\begin{eqnarray}\label{QLE}
\dot{a}&=&-i(\omega_{c}-i\kappa )a-i\lambda _{r}b-i\lambda _{\text{cr}}b^{\dag}+\sqrt{2\kappa}a_{\text{in}},\nonumber\\
\dot{b}&=&-i(\omega_{m}-i\gamma )b-iKb^{\dag}bb-i\lambda_{r}a-i\lambda_{\text{cr}}a^{\dag}+\sqrt{2\gamma }b_{\text{in}},
\end{eqnarray}
where $\gamma$ denotes the damping rate of the magnon mode, and $a_{\text{in}}$ ($b_{\text{in}}$) is the input noise operator related to the cavity mode (magnon mode), with zero mean value $\langle a_{\rm in} \rangle=\langle b_{\rm in} \rangle=0$.
Note that the cavity and magnon losses [corresponding to the dissipative terms $-i\kappa a$ and $-i\gamma b$ in Eq.~(\ref{QLE})] do not destroy the parity symmetry of the system, which has been verified in Ref.~\cite{Zhang21}. In the Markovian framework, the noise operators $a_{\text{in}}$ and $b_{\text{in}}$ satisfy the non-zero correlation functions ($o=a,\,b$)
\begin{eqnarray}\label{}
\langle o_{\rm in}^\dag(t) o_{\rm in}(t')\rangle&=& n_o\delta(t-t'),\nonumber\\
\langle o_{\rm in}(t) o_{\rm in}^\dag(t')\rangle&=& (n_o+1)\delta(t-t'),
\end{eqnarray}
where $n_a=(e^{\hbar \tilde{\omega}_c/k_B T}-1)^{-1}$ and $n_b=(e^{\hbar \tilde{\omega}_m/k_B T}-1)^{-1}$, with the Boltzmann constant $k_B$ and the bath temperature $T$, are the average thermal excitations in the thermal baths coupled to the cavity mode and the magnon mode, respectively.

\begin{figure}
\includegraphics[width=0.48\textwidth]{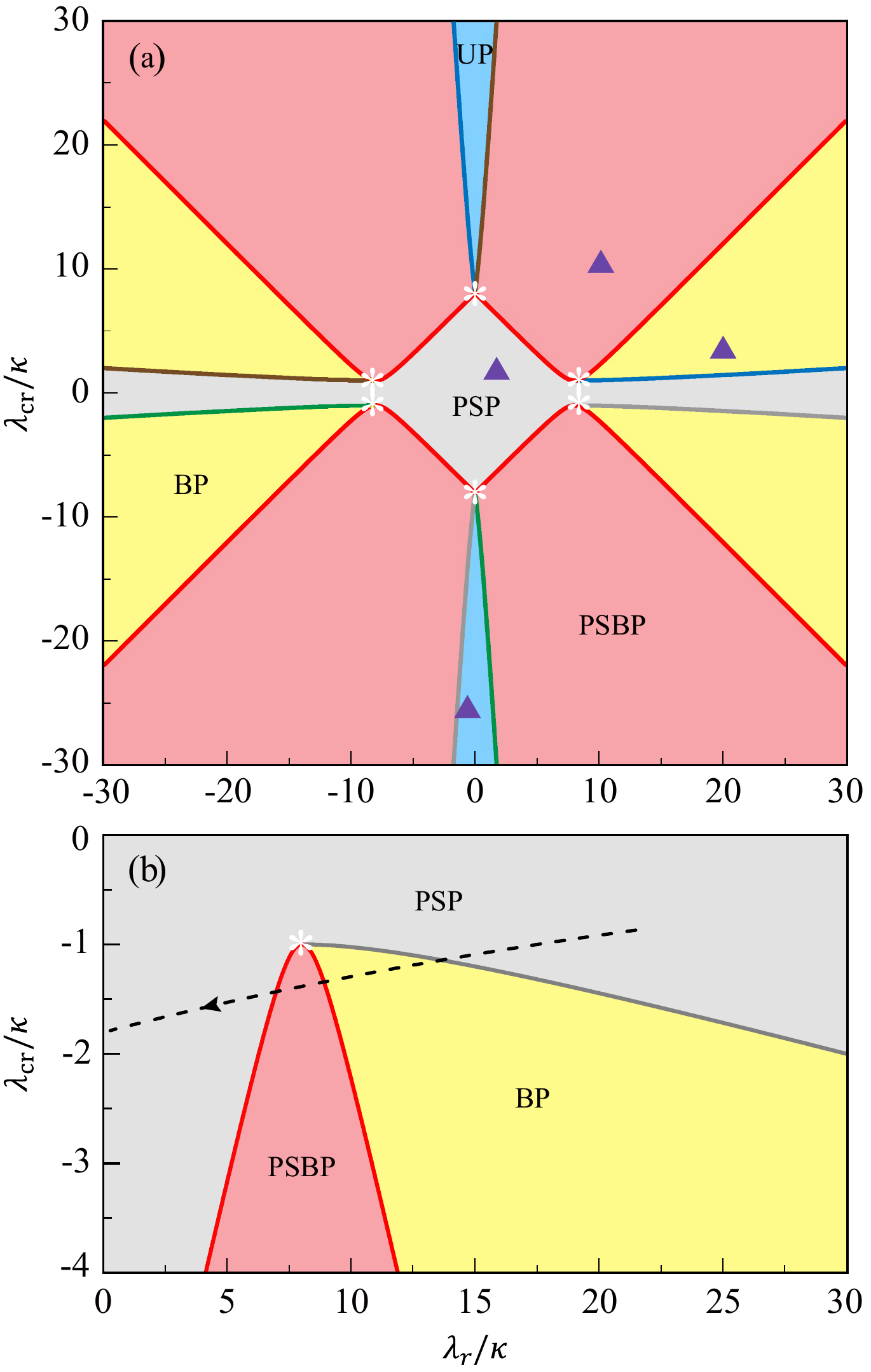}
\caption{(a) Steady-state phase diagram for the cavity magnonic system. Gray, red, yellow and blue regions represent the parity-symmetric phase (PSP), parity-symmetry-broken phase (PSBP), bistable phase (BP), and unstable phase (UP), respectively. The solid curves of different colors denote phase boundaries, with white asterisks marking the tricritical points and purple triangles indicating the specific positions chosen for the numerical simulations in Fig.~\ref{fig4}. The used parameters are the same as in Fig.~\ref{fig2}. (b) Enlarged view of a portion of (a), showing the region $0<\lambda_r/\kappa<30$ and $-4<\lambda_{\rm cr}/\kappa<0$. The black dashed curve with an arrow indicates the trajectory of $(\lambda_r,\,\lambda_{\rm cr})$ as $\xi$ increases from 2.05 to 2.4, starting at $(\lambda_r,\,\lambda_{\rm cr})=(21.47,-0.87)$ and ending at $(0.28,-1.79)$. This trajectory explicitly illustrates the phase transitions shown in Fig.~\ref{fig5}.}
\label{fig3}
\end{figure}

\begin{figure}
\includegraphics[width=0.48\textwidth]{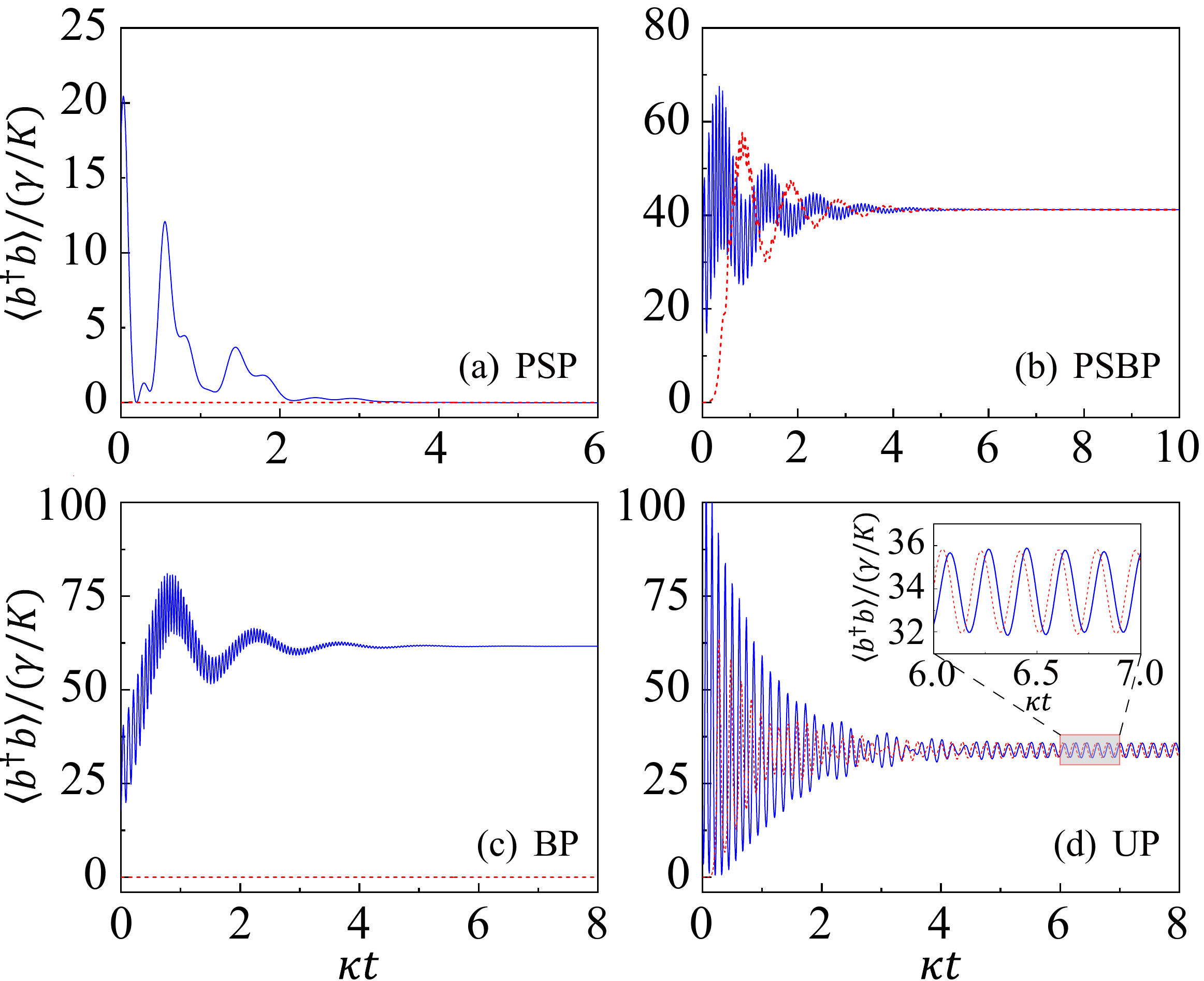}
\caption{(a)-(d) Temporal evolution of the scaled magnon number $\langle b^\dag b\rangle/(\gamma/K)$ at the points marked by purple triangles in Fig.~\ref{fig3}: (a) the parity-symmetric phase ($\lambda_{r}/\kappa=\lambda_{\text{cr}}/\kappa=2$), (b) the parity-symmetry-broken phase ($\lambda_{r}/\kappa=\lambda_{\text{cr}}/\kappa=10$), (c) the bistable phase ($\lambda_{r}/\kappa=20$, $\lambda_{\text{cr}}/\kappa=4$), and (d) the unstable phase ($\lambda_{r}/\kappa=-1$, $\lambda_{\text{cr}}/\kappa=-25$). The inset in (d) displays an enlarged view of the temporal evolution of $\langle b^\dag b\rangle/(\gamma/K)$ for the time interval $6 \leq \kappa t \leq 7$. In (a)-(d), the solid blue curves correspond to initial conditions $\langle a\rangle_{t=0}/\sqrt{\gamma/K}=-7.8-5.7i$ and $\langle b\rangle_{t=0}/\sqrt{\gamma/K}=3.1+2.9i$, while the dashed red curves represent $\langle a\rangle_{t=0}/\sqrt{\gamma/K}=-0.1-0.1i$ and $\langle b\rangle_{t=0}/\sqrt{\gamma/K}=0.1+0.1i$. Other parameters are the same as in Fig.~\ref{fig2}.}
\label{fig4}
\end{figure}

In order to linearize Eq.~(\ref{QLE}), we expand the operators $a$ and $b$ as the form of $a=\langle a\rangle+\delta a$ and $b=\langle b\rangle+\delta b$, where $\langle a\rangle$ ($\langle b\rangle$) represents the expectation value of the operator $a$ ($b$), and $\delta a$ ($\delta b$) is the corresponding quantum fluctuation with $\langle\delta a\rangle=\langle\delta b\rangle=0$. From Eq.~(\ref{QLE}), the equations of motion for the expectation values $\langle a\rangle$ and $\langle b\rangle$ are
\begin{eqnarray}\label{semiclassical}
\langle\dot{a}\rangle&=&-i(\omega_c-i\kappa)\langle a\rangle-i\lambda _{r}\langle b\rangle-i\lambda_{\text{cr}}\langle b^{\dag}\rangle,\nonumber\\
\langle\dot{b}\rangle&=&-i\left(\omega_m+K\langle b^{\dag}\rangle\langle b\rangle-i\gamma\right)\langle b\rangle-i\lambda_{r}\langle a\rangle-i\lambda_{\text{cr}}\langle a^{\dag}\rangle.
\end{eqnarray}
At the steady state (i.e., $\langle\dot{a}\rangle=\langle\dot{b}\rangle=0$), the above equations of motion can be solved analytically, and the solutions for $\langle b^{\dag}b\rangle$ are given by
\begin{eqnarray}\label{solutions}
\langle b^{\dag}b\rangle_{0}&=&0,\nonumber\\
\langle b^{\dag}b\rangle_{\pm}&=&\frac{1}{(\omega_{c}^{2}+\kappa^{2})K}
\Bigg\{[(\lambda_{r}^{2}+\lambda_{\text{cr}}^{2})\omega_{c}-(\omega_{c}^{2}+\kappa^{2})\omega_m]\nonumber\\
&&\pm\sqrt{(2\lambda_{r}\lambda_{\text{cr}}\omega_{c})^{2}-[(\omega_{c}^{2}+\kappa^{2})\gamma+(\lambda_{r}^{2}-\lambda _{\text{cr}}^{2})\kappa]^{2}}\Bigg\},~~~~~~
\end{eqnarray}
where the mean-field approximation $\langle b^{\dag}b\rangle\approx\langle b^{\dag}\rangle \langle b \rangle$ has been used. The trivial solution $\langle b^{\dag}b\rangle=0$ and the nontrivial solution $\langle b^{\dag}b\rangle=\langle b^{\dag}b\rangle_+$ remain stable within specific parameter regions (see Figs.~\ref{fig3} and \ref{fig4} and related discussions). A second nontrivial solution, $\langle b^{\dag}b\rangle=\langle b^{\dag}b\rangle_-$, is unstable throughout the parameter space and therefore excluded from our analysis. For $\langle b^{\dag}b\rangle_+$ to be physically meaningful, it must satisfy $\langle b^{\dag}b\rangle_+ > 0$. The phase boundaries in Fig.~\ref{fig3} are thus determined by solving this positivity condition, as given by Eqs.~(\ref{threshold1}) and (\ref{threshold2}).

After neglecting the high-order terms of the fluctuations, it follows from Eq.~(\ref{QLE}) that the quantum fluctuations $\delta a$ and $\delta b$ satisfy
\begin{eqnarray}\label{QFE}
\delta\dot{a}&=&-i(\omega_{c}-i\kappa)\delta a-i\lambda _{r}\delta b-i\lambda _{\text{cr}}\delta b^{\dag}+\sqrt{2\kappa}a_{\text{in}},\nonumber\\
\delta\dot{b}&=&-i(\omega'_{m}-i\gamma)\delta b-i\lambda_{r}\delta a-iF\delta b^{\dag}-i\lambda _{\text{cr}}\delta a^{\dag}+\sqrt{2\gamma}b_{\text{in}},~~~~~
\end{eqnarray}
where $\omega'_m=\omega_{m}+2K\langle b^{\dag}\rangle\langle b\rangle$ and $F=K{\langle b\rangle}^{2}$. For convenience, we introduce the quadrature operators $X_a=(\delta a^{\dag}+\delta a)/\sqrt{2}$, $Y_a=i(\delta a^{\dag}-\delta a)/\sqrt{2}$, $X_b=(\delta b^{\dag}+\delta b)/\sqrt{2}$, and $Y_b=i(\delta b^{\dag}-\delta b)/\sqrt{2}$.
Similarly, the input noise quadratures are defined by $X_{a}^{(\text{in})}=(a_{\text{in}}^{\dag}+ a_{\text{in}})/\sqrt{2}$, $Y_{a}^{(\text{in})}=i(a_{\text{in}}^{\dag}- a_{\text{in}})/\sqrt{2}$, $X_{b}^{(\text{in})}=(b_{\text{in}}^{\dag}+ b_{\text{in}})/\sqrt{2}$, and $Y_{b}^{(\text{in})}=i(b_{\text{in}}^{\dag}-b_{\text{in}})/\sqrt{2}$. Using these operators, the equations of motion for the quadrature fluctuations in Eq.~(\ref{QFE}) can be cast in the matrix form,
\begin{eqnarray}
\dot{\mathcal{O}} = \mathcal{U}~ \mathcal{O} + \mathcal{O}_{\text{in}},
\end{eqnarray}
where $\mathcal{O}=(X_{a}, Y_{a}, X_{b}, Y_{b})^{T}$ is the vector of quadrature components,
$\mathcal{O}_{\text{in}}=(\sqrt{2\kappa}\,X_{a}^{(\text{in})},\sqrt{2\kappa}\,Y_{a}^{(\text{in})},\sqrt{2\gamma}\,X_{b}^{(\text{in})},
\sqrt{2\gamma}\,Y_{b}^{(\text{in})})^{\text{T}}$ is the vector of noise quadratures, and the drift matrix $\mathcal{U}$ is given by
\begin{equation}\label{matrixU}
\begin{split}
\mathcal{U}=\left(
\begin{array}{cccc}
-\kappa & \omega_c & 0 & \lambda_{r}-\lambda_{\text{cr}} \\
-\omega_c & -\kappa & -\lambda_{r}-\lambda_{\text{cr}} & 0 \\
0 & \lambda_{r}-\lambda_{\text{cr}} & -\gamma+\text{Im}[F]  & \omega_{m}^{'}-\text{Re}[F] \\
 -\lambda_{r}-\lambda_{\text{cr}} & 0 & -\omega_{m}^{'}-\text{Re}[F] & -\gamma-\text{Im}[F]
\end{array}
\right) .
\end{split}
\end{equation}
The stability of a solution for $\langle b^\dag b\rangle$ from Eq.~(\ref{solutions}) is assessed by computing the eigenvalues of its corresponding matrix $\mathcal{U}$; stability requires all their real parts to be negative~\cite{Gradshteyn80}.

\section{Quantum phase transition}\label{QPT}

By analyzing the stability of all solutions in Eq.~(\ref{solutions}), we plot the steady-state phase diagram of the cavity magnonic system as functions of the coupling strengths $\lambda_r/\kappa$ and $\lambda_{\rm cr}/\kappa$ in Fig.~\ref{fig3}(a). There exists four different regions in the phase diagram, which represents the parity-symmetric phase (gray area), parity-symmetry-broken phase (red area), bistable phase (yellow area), and unstable phase (blue area), respectively. The phase diagram features a continuous (second-order) boundary separating the parity-symmetric phase from the parity-symmetry-broken phase, and a discontinuous (first-order) boundary separating the parity-symmetric phase from the bistable phase (i.e., the parity-symmetry-broken phase in the bistable region~\cite{Zhang21}). Phase boundaries are depicted by solid curves of various colors, and tricritical points are indicated by white asterisks. These features follow from the condition $\langle b^{\dagger}b\rangle_+ > 0$. The tricritical points are located at $(\pm 8, \pm 1)$ and $(0,\pm \sqrt{65})$, where three different phases meet. The phase boundaries are characterized by two sets of critical parameters. The first set is given by
\begin{equation}\label{threshold1}
\lambda_{\text{cr1}}^{(\pm,\pm)} = \pm\sqrt{\frac{\omega_{c}^{2}+\kappa^{2}}{\kappa^{2}}(\kappa\gamma + \lambda_{r}^{2})} \pm \frac{\omega_{c}}{\kappa}\lambda_r,
\end{equation}
where $\lambda_{\text{cr1}}^{(+,+)}$, $\lambda_{\text{cr1}}^{(+,-)}$, $\lambda_{\text{cr1}}^{(-,+)}$ and $\lambda_{\text{cr1}}^{(-,-)}$ correspond to the brown, blue, gray and green solid curves, respectively. The second set satisfies
\begin{equation}\label{threshold2}
\lambda_{\text{cr2}}^{\pm} = \pm\sqrt{\lambda_{r}^{2} + \omega_c\omega_m + \kappa\gamma - \sqrt{4\omega_c\omega_m\lambda_{r}^{2} - (\omega_{c}\gamma-\omega_{m}\kappa)^{2}}},
\end{equation}
which corresponds to the red solid curves. In the parity-symmetric phase, the parity symmetry of the system is conserved, and the solution $\langle b^{\dag}b\rangle=0$ is stable, indicating no macroscopic magnon occupation in the steady state [see Fig.~\ref{fig4}(a)]. When the parity symmetry is broken (corresponding to the parity-symmetry-broken phase), the system has macroscopic magnon excitations with $\langle b^{\dag}b\rangle=\langle b^{\dag}b\rangle_{+}$ ($>0$) [see Fig.~\ref{fig4}(b)]. In the parity-symmetry-broken phase, the system exhibits two degenerate steady states, reading $\langle b\rangle=\pm\sqrt{\langle b^{\dag}b\rangle_{+}}\, e^{i\theta/2}$, with $e^{i\theta}=(\sqrt{\Sigma^2-\Gamma^2}+i\Gamma)/\Sigma$, $\Sigma=2\lambda_r\lambda_{\rm cr}\omega_c$, and $\Gamma=(\omega_c^2+\kappa^2)\gamma+(\lambda_r^2-\lambda_{\rm cr}^2)\kappa$. The degeneracy is guaranteed by the parity symmetry of the system.
Under the transformation $(\langle a\rangle,\,\langle b\rangle)\mapsto (-\langle a\rangle,\,-\langle b\rangle)$, the semiclassical equation of motion in Eq.~(\ref{semiclassical}) remains invariant. As in the superradiant phase of the open Dicke model~\cite{Baumann11,Ning24}, the environment drives the system in this phase to spontaneously select one of the two steady states. Different from parity-symmetric and parity-symmetry-broken phases, both solutions $\langle b^{\dag}b\rangle=0$ and $\langle b^{\dag}b\rangle=\langle b^{\dag}b\rangle_{+}$ are stable in the bistable region, where the initial state of the system determines whether it will evolve into the steady state with or without macroscopic magnon occupation [see Fig.~\ref{fig4}(c)]. Remarkably, the system displays the dynamical instability in the unstable phase, with all three solutions in Eq.~(\ref{solutions}) becoming unstable [see Fig.~\ref{fig4}(d)].

\begin{figure}
\includegraphics[width=0.48\textwidth]{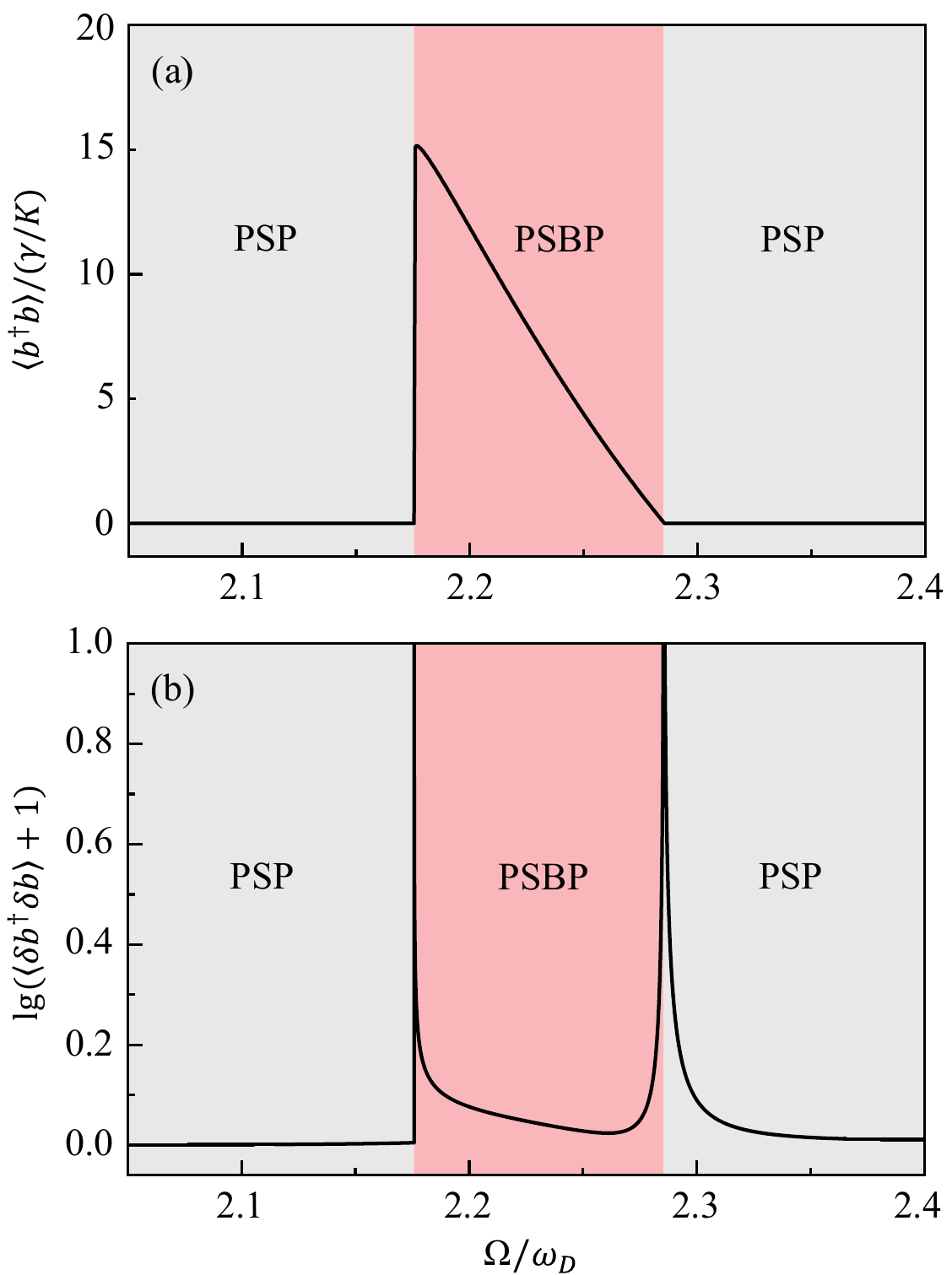}
\caption{(a) The scaled magnon number $\langle b^{\dag}b\rangle /(\gamma / K)$ and (b) the magnon number fluctuation $\lg(\langle \delta b^{\dag}\delta b\rangle + 1)$ versus the reduced drive amplitude $\Omega/\omega_{D}$, with $n_a=n_b=0$. The abbreviations PSP and PSBP denote the parity-symmetric phase and parity-symmetry-broken phase, respectively. Other parameters are the same as in Fig.~\ref{fig2}(b).}
\label{fig5}
\end{figure}

In Fig.~\ref{fig5}(a), we display the scaled magnon number $\langle b^{\dag}b\rangle/(\gamma/K)$ as a function of the reduced drive amplitude $\Omega/\omega_{D}$. As $\Omega/\omega_{D}$ increases from 2.05 to 2.4,the system follows the black dashed trajectory in the $(\lambda_r, \lambda_{\text{cr}})$ plane shown in Fig.~\ref{fig3}(b); the corresponding $\lambda_r(\Omega/\omega_{D})$ and $\lambda_{\text{cr}}(\Omega/\omega_{D})$ are shown in Fig.~\ref{fig2}(b). For $\Omega/\omega_{D} < 2.176$, the cavity magnonic system remains in the parity-symmetric phase, characterized by $\langle b^{\dag}b\rangle/(\gamma/K) = 0$. Near $\Omega/\omega_{D} = 2.176$, the scaled magnon number $\langle b^\dag b\rangle/(\gamma/K)$ exhibits a discontinuous jump from zero to a finite value $\langle b^\dag b\rangle_+/(\gamma/K)$ ($\neq 0$), which marks that the system enters the parity-symmetry-broken phase. This behavior signals a first-order SPT accompanied by spontaneous parity symmetry breaking. With further increase of $\Omega/\omega_{D}$, $\langle b^\dag b\rangle/(\gamma/K)$ decreases continuously until, at $\Omega/\omega_{D} = 2.285$, the system undergoes a second-order SPT and reenters the parity-symmetric phase, where $\langle b^\dag b\rangle/(\gamma/K)$ become zero again.

Furthermore, we investigate the magnon number fluctuation $\langle \delta b^{\dag}\delta b\rangle$ during the quantum phase transition. To compute this quantity, we introduce the time-dependent covariance matrix $\mathcal{V}(t)$, with matrix elements defined as  $\mathcal{V}_{\rm ij}(t)=\langle \mathcal{O}_i(t)\mathcal{O}_j(t')+\mathcal{O}_j(t')\mathcal{O}_i(t)\rangle/2$ for $i,j = 1,2,3,4$. In the steady state, the stationary covariance matrix $\mathcal{V}=\mathcal{V}(t=+\infty)$ satisfies the Lyapunov equation~\cite{Parks93,Vitali07}
\begin{equation}
\mathcal{U}\mathcal{V}+\mathcal{V}\mathcal{U}^{T}=-\mathcal{D},
\end{equation}
where the diffusion matrix $\mathcal{D} = \mathrm{diag}[(2n_a+1)\kappa, (2n_a+1)\kappa, (2n_b+1)\gamma, (2n_b+1)\gamma]$ is defined via $D_{\rm ij}\delta(t-t')=\langle O_{\text{in},i}(t) O_{\text{in},j}(t')+ O_{\text{in},j}(t') O_{\text{in},i}(t)\rangle/2$, and the drift matrix $\mathcal{U}$ is given in Eq.~(\ref{matrixU}). Solving the Lyapunov equation yields the magnon number fluctuation as $\langle \delta b^{\dag} \delta b \rangle = \big[ (V_{33} + V_{44}) - 1 \big] / 2$~\cite{Zhu20}. For clarity, we plot $\lg \big( \langle \delta b^\dag \delta b \rangle + 1 \big)$ as a function of the reduced drive amplitude $\Omega/\omega_{D}$ in Fig.~\ref{fig5}(b). The fluctuation $\lg \big( \langle \delta b^\dag \delta b \rangle + 1 \big)$ diverges near the critical points $\Omega/\omega_{D} = 2.176$ and $\Omega/\omega_{D} = 2.285$, but tends to zero away from these values.

\section{Discussions and Conclusions}

In the cavity-magnon system, both rotating- and counter-rotating terms are intrinsically present regardless of the cavity-magnon coupling strength~\cite{Rameshti22,Yuan22}. References~\cite{Zhang21,Qin22,Xu24} drop the counter-rotating terms within the rotating-wave approximation, justified by their negligible contribution under parametric drive. In the present scheme, however, this approximation is no longer applicable, since the Floquet drive induces sidebands that prevent such a treatment [cf.~Eq.(\ref{heff1}) and related discussions]. Although we retain the counter-rotating terms, the system remains in the strong-coupling regime (i.e., $\{\kappa,\gamma\} \ll g_m \ll \{\tilde{\omega}_c,\,\tilde{\omega}_m\}$), which is readily accessible in experiments~\cite{Rameshti22,Yuan22}. In addition to the counter-rotating terms, the magnon Kerr effect plays a crucial role in our scheme. Without the Kerr nonlinearity (i.e., $K=0$), we find that the semiclassical equation of motion in Eq.~(\ref{semiclassical}) admits only the trivial steady-state solution $\langle b^\dag b\rangle_0=0$. In that case, even with the counter-rotating terms present, the system supports only the parity-symmetric phase and an unstable phase; no phase transition occurs. The SPT in the standard Dicke model (equivalent to two nonlinearly coupled harmonic oscillators) originates from this nonlinear coupling~\cite{Emary03PRL,Emary03PRE}. Here, the linear cavity-magnon coupling (including both rotating and counter-rotating terms) and the magnon Kerr effect together play a role analogous to the nonlinear coupling in the standard Dicke model.

In summary, we present a Floquet-modulation approach to engineer the SPT in cavity magnonics. By periodically modulating the magnon frequency, the effective frequencies of the cavity and magnon modes, as well as their coupling strength, can be flexibly tuned. The steady-state analysis reveals a rich phase diagram of the system, including parity-symmetric, parity-symmetry-broken, bistable, and unstable phases. With increasing Floquet-field strength, the system exhibits a first-order phase transition from the parity-symmetric phase to the parity-symmetry-broken phase, followed by a second-order phase transition that returns the system to the parity-symmetric phase again. Enhanced magnon number fluctuations near the critical points further confirm the occurrence of the phase transitions. Our work provides an alternative approach for exploring the cavity-magnon SPT.

\section*{Acknowledgments}

This work is supported by the National Natural Science Foundation of China (Grants No. 12674436, No. 12205069, No. 12504343, and No. 12404403), the HZNU scientific research and innovation team project (Grant No. TD2025003), the Hangzhou Leading Youth Innovation and Entrepreneurship Team project (Grant No. TD2024005), and the Zhejiang Provincial Natural Science Foundation of China (Grants No. LQN25A040018 and No. LQN25A040019). Y.J.X. is supported by the Natural Science Foundation for Distinguished Young Scholars of the Higher Education Institutions of Anhui Province (Grant No. 2022AH020097).

\end{document}